\documentclass[twocolumn, Journal]{IEEEtran}

\IEEEoverridecommandlockouts

\usepackage{color}
\usepackage{bm}
\usepackage{graphicx}
\usepackage{amsmath}
\usepackage{amssymb}
\usepackage{algorithm}
\usepackage{algorithmic}
\usepackage{amsmath}
\usepackage{multirow}
\usepackage{booktabs}
\usepackage{array}
\usepackage{amsthm}
\usepackage{lipsum}
\usepackage{enumerate}
\usepackage{stfloats}
\usepackage{subfigure}
\usepackage{cases}
\usepackage{diagbox}

\newtheorem{remark}{Remark}

\newcommand{\non}{\nonumber}
\title{Rate-Splitting Multiple Access for 6G -- Part III: Interplay with Reconfigurable Intelligent Surfaces}

\author{Hongyu Li, \IEEEmembership{Student Member, IEEE}, Yijie Mao, \IEEEmembership{Member, IEEE}, Onur Dizdar, \IEEEmembership{Member, IEEE}, \\and Bruno Clerckx, \IEEEmembership{Fellow, IEEE}\\
\textit{(Invited Paper)}
\thanks{Manuscript received May 4, 2022; revised June 21, 2022; accepted July 12, 2022. The associated editor coordinating the review of this letter and approving it for publication was Dr. Alessio Zappone.}
\thanks{H. Li, O. Dizdar, and B. Clerckx are with Department of Electrical and Electronic Engineering, Imperial College London, London SW7 2AZ, U.K. (email: \{c.li21, o.dizdar, b.clerckx\}@imperial.ac.uk).}
\thanks{Y. Mao is with the School of Information Science and Technology, ShanghaiTech University, Shanghai 201210, China (email: maoyj@shanghaitech.edu.cn).}}

\begin{document}

\maketitle

\begin{abstract}
This letter is the third part of a three-part tutorial that focuses on rate-splitting multiple access (RSMA) for 6G. As Part III of the tutorial, this letter provides an overview of integrating RSMA and reconfigurable intelligent surface (RIS). We first introduce two potential PHY layer techniques, namely, RSMA and RIS, including the need for integrating RSMA with RIS and how they could help each other. Next, we provide a general model of an RIS-aided RSMA system and summarize some key performance metrics. Then, we discuss the major advantages of RIS-aided RSMA networks, and illustrate the rate region of RIS-aided RSMA for both perfect and imperfect channel conditions. Finally, we summarize the research challenges and open problems for RIS-aided RSMA systems. In conclusion, RSMA is a promising technology for next generation multiple access (NGMA) and future networks such as 6G and beyond.      
\end{abstract}


\maketitle

\vspace{-0.3 cm}

\section{Introduction}

%\subsection{Background}

%%% 6G Background
With the deployment of 5G New Radio (NR) networks, research focuses of both academia and industry have shifted to the next generation of wireless communications, namely, 6G. 
6G is expected to improve existing techniques and/or develop novel techniques at the infrastructure level, the spectrum level, and the protocol/algorithmic level to provide better services and realize full-coverage connectivity \cite{Tataria}, \cite{Rajatheva}.  
Among various enabling techniques, rate-splitting multiple access (RSMA) \cite{Lina Mao}, \cite{Lina Mao 2018} at the protocol/algorithmic level and reconfigurable intelligent surface (RIS) \cite{Di Renzo 2020} at the infrastructure level are two important techniques that have gained immense attention.

%%% RSMA Overview
RSMA is envisioned as a promising non-orthogonal transmission strategy for beyond 5G wireless communications \cite{Lina Mao}. The main principle of RSMA is to split (and combine) user messages into common and private messages, which are independently encoded into common streams shared by multiple users and private ones intended for different users.
Such scheme enables a flexible interference management by decoding part of the interference and treating part of the interference as noise \cite{Lina Mao 2018}. Existing research has demonstrated various advantages of RSMA, such as \textit{i)} performance enhancement, e.g., spectral efficiency \cite{Lina Mao 2018}, \textit{ii)} robustness to channel state information (CSI) conditions \cite{Anup} and mobility \cite{Onur Dizdar}, and \textit{iii)} the superiority in reducing latency \cite{Onur Dizdar}, compared to other multiple access techniques, such as space division multiple access (SDMA), power-domain non-orthogonal multiple access (NOMA)\footnote{In the rest of this letter, we use NOMA to represent power-domain NOMA for simplicity.}, and orthogonal multiple access (OMA).

% \subsubsection{Disadvantages}
% \begin{itemize}
%   \item Relatively High Signal Processing Complexity (for generalized RSMA):
% \end{itemize} 

%%% RIS Overview
Meanwhile, RIS is deemed as another revolutionary technique for future wireless networks \cite{Di Renzo 2020}. 
RIS systems benefit from two-dimensional planar surfaces, usually consisting of large numbers of passive tuneable elements, to enhance the performance of multi-antenna systems.
There are twofold advantages of RIS: \textit{i)} each element of an RIS is realized by cost- and energy-efficient passive devices, e.g., positive intrinsic negative (PIN) diodes, so that RISs can be employed in practical systems with extremely low power consumption; \textit{ii)} each element of an RIS is reconfigurable, so that, RISs are able to adaptively shape transmission  environments and enhance transmission quality. These two advantages have motivated researchers to combine RISs with various schemes, e.g., NOMA \cite{B Zheng}, wireless power transfer (WPT) \cite{Kris}, and integrated sensing and communication (ISAC) \cite{RSP}.

% \subsubsection{Disadvantages}

% \begin{itemize}
%   \item Difficult to acquire accurate CSI:
%   \item Not flexible for user scheduling: 
%   \item Weak for mobility and latency: 
% \end{itemize}

%\subsection{The Integration of RSMA and RIS}

%%% RIS + RSMA
Inspired by the appealing advantages of RIS and RSMA as well as the performance benefits of synergizing RIS and other techniques \cite{B Zheng}-\cite{RSP}, some recent works \cite{Z Yang}-\cite{A Jolly} have investigated the interplay between RSMA and RIS, and shown their benefits in terms of energy efficiency maximization \cite{Z Yang}, \cite{K Weinberger Journal}, outage probability analysis \cite{A Bansal}, \cite{A Bansal Open J}, transmit power minimization \cite{K Weinberger}, max-min fairness \cite{H Fu}, and spectral efficiency maximization \cite{T Fang}, \cite{A Jolly}.
However, none of the above works explain the fundamental motivation to integrate RSMA with RIS.   
Before we start working on RIS-aided RSMA, there is an important question worth discussing: Why we should consider the integration of RSMA and RIS? In the following we will answer this question from a broad perspective by discussing the similarities between RSMA and RIS, and how they can complement each other. 

% \subsection{RIS-Aided RSMA}

% Why RIS + RSMA? (A table of literature review on RIS+RSMA)

\textit{Similarities:} There are two similarities between RSMA and RIS. On one hand, both are suitable for (and have benefits in) multi-antenna systems. On the other hand, both aim at designing (either active or passive) beamforming to achieve better performance.  

\textit{Complementarities:} RSMA and RIS complement each other in several aspects. From the RIS side, since RISs are essentially passive devices, they cannot transmit signals by themselves like active devices, e.g., base stations. Given this fact, RISs should be deployed in wireless communication networks as an ``assistant'' to establish a new propagation path between the transmitter and the receiver so as to enhance communication quality. This can be considered as the fundamental reason for why we need to consider the interplay between RIS and other techniques, such as RSMA. On the other hand, accurate and instantaneous CSI acquisition is difficult in RIS-aided systems since RIS elements are passive \cite{Di Renzo 2020}. As RSMA is robust to imperfect CSI conditions \cite{Anup} and high mobility scenarios \cite{Onur Dizdar}, it is possible to ease the stringent requirement of CSI in RIS-aided RSMA networks while maintaining good performance. 
From the RSMA side, with the aid of RISs, low-complexity RSMA schemes such as 1-layer RSMA, are able to achieve enhanced performance \cite{A Bansal}, which can effectively avoid the use of RSMA schemes with complex architectures, such as generalized RSMA \cite{Lina Mao}, and thus reduce receiver/encoding/signalling complexities.

Based on the above discussion, we can deduce that integrating RSMA and RIS is a win-win solution for both techniques. In the rest of this letter, we will discuss the modeling of RIS-aided RSMA networks and show the advantages of integrating RSMA with RIS. In Section II, we develop a general model for RIS-aided RSMA communication systems, which covers all existing RIS-aided RSMA schemes, and summarize commonly used performance metrics. Then in Section III, we discuss the advantages of RIS-aided RSMA networks and provide simulation results. Finally, in Section IV, we give a brief outlook of possible open problems for RIS-aided RSMA.

% \textit{Notation:} Vectors and matrices are denoted by bold lowercase and uppercase letters, respectively. $(\cdot)^T$ and $(\cdot)^H$ denote the transpose and Hermitian transpose operation, respectively. $\mathcal{CN}(0, \sigma^2)$ denotes the Circularly Symmetric
% Complex Gaussian distribution with zero mean and variance $\sigma^2$. $\mathsf{diag}(\cdot)$ and $\mathsf{blkdiag}(\cdot)$ denote the diagonal matrix and the block-diagonal matrix, respectively. 

\section{RIS-Aided RSMA Transmission Networks}

\begin{figure}
  \centering
  \includegraphics[height=2 in]{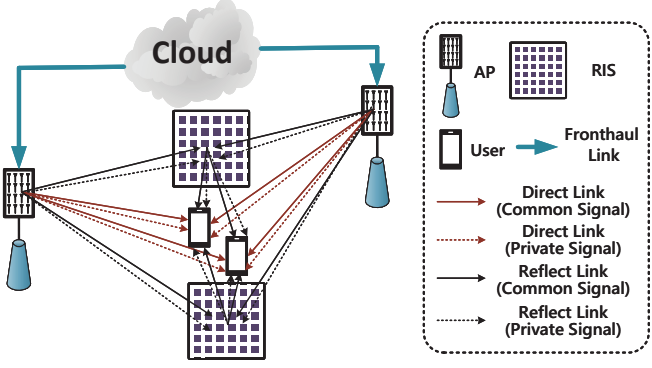}
  \caption{A paradigm of a multi-RIS-aided RSMA system.}
  \label{fig:syst_mod} \vspace{-0.5 cm}
\end{figure}

Consider a general multi-RIS-aided cell-free downlink\footnote{Although we focus on the modeling of downlink RIS-aided RSMA in this work, the modeling of RIS introduced in this section is suitable for both uplink and downlink scenarios since the essence of RIS is to modify the channel between the transmitter and the receiver. Thus, the formulation of uplink RIS-aided RSMA is straightforward based on \cite{Lina Mao} and Part I of this tutorial with simple modifications of channels as in equation (1).} system with $N$ access points (APs) as shown in Fig. \ref{fig:syst_mod}. A central cloud connected to all APs processes baseband signals and performs tasks, such as splitting user messages, encoding messages into signals, designing transmit precoders, and sending them to APs. Each AP equipped with $N_\mathrm{t}$ antennas simultaneously communicates with $K$ single-antenna users. $L$ RISs, each of which has $M$ elements indexed by $\mathcal{M} = \{1, \ldots, M\}$, are distributed in the network to support the communications between APs and users. The sets of the indices of APs, RISs, and users are defined as $\mathcal{N} = \{1,\ldots,N\}$, $\mathcal{L} = \{1, \ldots, L\}$, and $\mathcal{K} = \{1, \ldots, K\}$, respectively.

Let $\mathbf{x} = [\mathbf{x}_1^T, \ldots, \mathbf{x}_N^T]^T \in \mathbb{C}^{NN_\mathrm{t}}$ be the vector of the transmit signal , where $\mathbf{x}_n \in \mathbb{C}^{N_\mathrm{t}}$ denotes the transmit signal vector at AP-$n$ and satisfies a power constraint $\mathbb{E}\{\mathbf{x}_n^H\mathbf{x}_n\} \le P_n$, with $P_n$ representing the maximum transmit power at AP-$n$, $\forall n \in \mathcal{N}$.  Then the signal received at each user is given by
\begin{equation}
  y_k = \mathbf{h}_{\mathrm{d},k}^H\mathbf{x} + \mathbf{h}_{\mathrm{r},k}^H\mathbf{\Phi}\mathbf{G}\mathbf{x} + z_k, \forall k \in \mathcal{K},
\end{equation}
where $\mathbf{h}_{\mathrm{d},k} = [\mathbf{h}_{\mathrm{d},1,k}^H, \ldots, \mathbf{h}_{\mathrm{d},N,k}^H]^H \in \mathbb{C}^{NN_\mathrm{t}}$ denotes the aggregated direct channel between APs and user-$k$, and $\mathbf{h}_{\mathrm{d},n,k} \in \mathbb{C}^{N_\mathrm{t}}$ is the direct channel between AP-$n$ and user-$k$, $\forall n \in \mathcal{N}$, $\forall k \in \mathcal{K}$.
Similarly, $\mathbf{h}_{\mathrm{r},k} = [\mathbf{h}_{\mathrm{r},1,k}^H, \ldots, \mathbf{h}_{\mathrm{r},L,k}^H]^H \in \mathbb{C}^{LM}$ denotes the aggregated channel between RISs and user-$k$, where $\mathbf{h}_{\mathrm{r},l,k} \in \mathbb{C}^{M}$ is the channel between RIS-$l$ and user-$k$, $\forall l \in \mathcal{L}$, $\forall k\in \mathcal{K}$. 
$\mathbf{G} = [\mathbf{G}_1^H, \ldots, \mathbf{G}_L^H]^H \in \mathbb{C}^{LM \times NN_\mathrm{t}}$ with $\mathbf{G}_l = [\mathbf{G}_{l,1}, \ldots, \mathbf{G}_{l,N}] \in \mathbb{C}^{M \times NN_\mathrm{t}}$ denotes the aggregated channel between BSs and RISs, where $\mathbf{G}_{l,n} \in \mathbb{C}^{M \times N_\mathrm{t}}$ represents the channel between BS-$n$ and RIS-$l$, $\forall n \in \mathcal{N}$, $\forall l \in \mathcal{L}$. 
In addition, the block diagonal matrix $\mathbf{\Phi} = \mathsf{blkdiag}(\mathbf{\Phi}_1, \ldots, \mathbf{\Phi}_L)$ is defined as the overall RIS matrix, where $\mathbf{\Phi}_l \in \mathbb{C}^{M\times M}$ represents the coefficients of RIS-$l$, $\forall i \in \mathcal{L}$.
Finally, $z_k \sim \mathcal{CN}(0,\sigma_z^2)$, $\forall k \in \mathcal{K}$ is the additive white Gaussian noise (AWGN) at each user. It is worth noting that both RSMA and RIS have various architectures, thereby generating different transmit signal $\mathbf{x}$ and the problem design. In the following we will provide a brief summary of RSMA and RIS architectures.

\subsection{Architecture for RSMA} 

The main idea of RSMA is to split user messages into multiple sub-messages so as to achieve more flexible interference management compared to other multiple access techniques, such as SDMA, NOMA, and OMA. Different RSMA architectures have different ways of message splitting \cite{Lina Mao}. Here we introduce two most commonly used RSMA architectures, namely, 1-layer RS \cite{Lina Mao 2018} and 2-layer hierarchical RS (HRS) \cite{Lina Mao 2018}.

\subsubsection{1-layer RS} When considering a simple 1-layer RSMA scheme, each AP transmits superimposed streams, which contain one common stream $s_{\mathrm{c}}$ and $K$ private streams $s_{\mathrm{p},k}$, $\forall k \in \mathcal{K}$ to $K$ users. These streams are linearly precoded by a precoding matrix $\mathbf{P} = [\mathbf{p}_{\mathrm{c}}, \mathbf{p}_{\mathrm{p},1}, \ldots, \mathbf{p}_{\mathrm{p},K}] \in \mathbb{C}^{NN_\mathrm{t}\times (K + 1)}$ with precoders for common streams $\mathbf{p}_{\mathrm{c}} = [\mathbf{p}_{\mathrm{c},1}^T,\ldots, \mathbf{p}_{\mathrm{c},N}^T]^T \in \mathbb{C}^{NN_\mathrm{t}}$, $\mathbf{p}_{\mathrm{c},n} \in \mathbb{C}^{N_\mathrm{t}}$, and precoders for private streams 
$\mathbf{p}_{\mathrm{p},k} = [\mathbf{p}_{\mathrm{p},1,k}^T,\ldots, \mathbf{p}_{\mathrm{p},N,k}^T]^T \in \mathbb{C}^{NN_\mathrm{t}}$, $\mathbf{p}_{\mathrm{p},n,k} \in \mathbb{C}^{N_\mathrm{t}}$, $\forall n \in \mathcal{N}$, $\forall k \in \mathcal{K}$, such that the transmit signal  can be written as
\begin{equation}
  \mathbf{x} = \mathbf{p}_{\mathrm{c}}s_\mathrm{c} + \sum_{k \in \mathcal{K}}\mathbf{p}_{\mathrm{p},k}s_{\mathrm{p},k}.
\end{equation}

At user sides, each user first decodes the common stream by considering all private streams as noise. Therefore, the instantaneous rate for decoding the common stream can be calculated as 
\begin{equation}
  R_{\mathrm{c},k} = \log_2\left(1 + \frac{|\mathbf{h}_k^H\mathbf{p}_\mathrm{c}|^2}{\sum_{q \in \mathcal{K}} |\mathbf{h}_k^H\mathbf{p}_{\mathrm{p},q}|^2 + \sigma_z^2}\right), \forall k \in \mathcal{K},
\end{equation}
where $\mathbf{h}_k = (\mathbf{h}_{\mathrm{d},k}^H + \mathbf{h}_{\mathrm{r},k}^H\mathbf{\Phi}\mathbf{G})^H$, $\forall k \in \mathcal{K}$. In order to guarantee the successful decoding of the common message by all users, the common rate cannot exceed $R_\mathrm{c} = \min_{k \in \mathcal{K}} R_{\mathrm{c},k}$. After decoding the common stream, each user performs successive interference cancellation (SIC) to remove the common stream, and then decodes its own private stream, which yields the following private rate:
\begin{equation}
  R_{\mathrm{p},k} = \log_2\left(1 + \frac{|\mathbf{h}_k^H\mathbf{p}_{\mathrm{p},k}|^2}{\sum_{q \in \mathcal{K}, q \ne k} |\mathbf{h}_k^H\mathbf{p}_{\mathrm{p},q}|^2 + \sigma_z^2}\right), \forall k \in \mathcal{K}.
\end{equation}
After decoding the intended common and private streams, each user can reconstruct its original message by combining common and private sub-messages decoded from corresponding streams \cite{Lina Mao 2018}.

\subsubsection{2-layer HRS} When 2-layer HRS is considered, $K$ users are grouped into $G$ groups indexed by $\mathcal{G} = \{1, \ldots, G\}$, each of which contains $K_g$ ($\sum_{g \in \mathcal{G}} K_g = K$) users with an index set $\mathcal{K}_g$, $\cup_{g \in \mathcal{G}}\mathcal{K}_g = \mathcal{K}$. In this case, APs transmit superimposed streams containing one inter-group common stream $s_{\mathrm{c}1}$, $G$ inner-group common streams $s_{\mathrm{c}2,g}$, $\forall g \in \mathcal{G}$, and $K$ private streams $s_{\mathrm{p},k}$, $\forall k \in \mathcal{K}$, to $K$ users.
These streams are linearly precoded by $\mathbf{P} = [\mathbf{p}_{\mathrm{c}1}, \mathbf{p}_{\mathrm{c}2,1}, \mathbf{p}_{\mathrm{c}2,G}, \mathbf{p}_{\mathrm{p},1}, \ldots, \mathbf{p}_{\mathrm{p},K}] \in \mathbb{C}^{NN_\mathrm{t}\times (K + G + 1)}$ with precoders for inter-group common streams $\mathbf{p}_{\mathrm{c}1} = [\mathbf{p}_{\mathrm{c}1,1}^T,\ldots, \mathbf{p}_{\mathrm{c}1,N}^T]^T \in \mathbb{C}^{NN_\mathrm{t}}$, $\mathbf{p}_{\mathrm{c}1,n} \in \mathbb{C}^{N_\mathrm{t}}$, precoders for inner-group common streams $\mathbf{p}_{\mathrm{c}2,g} = [\mathbf{p}_{\mathrm{c}2,1,g}^T,\ldots, \mathbf{p}_{\mathrm{c}2,N,g}^T]^T \in \mathbb{C}^{NN_\mathrm{t}}$, $\mathbf{p}_{\mathrm{c}2,n,g} \in \mathbb{C}^{N_\mathrm{t}}$, and precoders for private streams 
$\mathbf{p}_{\mathrm{p},k} = [\mathbf{p}_{\mathrm{p},1,k}^T,\ldots, \mathbf{p}_{\mathrm{p},N,k}^T]^T \in \mathbb{C}^{NN_\mathrm{t}}$, $\mathbf{p}_{\mathrm{p},n,k} \in \mathbb{C}^{N_\mathrm{t}}$, $\forall n \in \mathcal{N}$, $\forall g \in \mathcal{G}$, $\forall k \in \mathcal{K}$. Therefore, the overall transmit signal vector is given as follows
\begin{equation}
  \mathbf{x} = \mathbf{p}_{\mathrm{c}1}s_{\mathrm{c}1} + \sum_{g \in \mathcal{G}} \mathbf{p}_{\mathrm{c}2,g}s_{\mathrm{c}2,g} + \sum_{k \in \mathcal{K}}\mathbf{p}_{\mathrm{p},k}s_{\mathrm{p},k}.
\end{equation}
Each user performs two layers of SIC to successively decode $s_{\mathrm{c}1}$, $s_{\mathrm{c}2,g}$, $\forall g \in \mathcal{G}$, and $s_{\mathrm{p},k}$, $\forall k \in \mathcal{K}$. Therefore, instantaneous rates for decoding inter-group common stream, inner-group common stream, and private stream are respectively given by $R_{\mathrm{c}1,k} = \log_2(1 + \gamma_{\mathrm{c}1,k})$, $R_{\mathrm{c}2,g,k} = \log_2(1 + \gamma_{\mathrm{c}2,g,k})$, and $R_{\mathrm{p},k} = \log_2(1 + \gamma_{\mathrm{p},k})$, $\forall k \in \mathcal{K}$, $\forall g \in \mathcal{G}$, where
\begin{equation}
  \non
    \gamma_{\mathrm{c}1,k} = \frac{|\mathbf{h}_k^H\mathbf{p}_{\mathrm{c}1}|^2}{\sum_{g \in \mathcal{G}}|\mathbf{h}_k^H\mathbf{p}_{\mathrm{c}2,g}|^2 + \sum_{q \in \mathcal{K}}|\mathbf{h}_k^H\mathbf{p}_{\mathrm{p},q}|^2 + \sigma_z^2}, 
\end{equation}
\begin{equation}
  \non
  \gamma_{\mathrm{c}2,g,k} = \frac{|\mathbf{h}_k^H\mathbf{p}_{\mathrm{c}2,g}|^2}{\sum_{\substack{o \in \mathcal{G}\\ o \ne g}}|\mathbf{h}_k^H\mathbf{p}_{\mathrm{c}2,o}|^2 + \sum_{q \in \mathcal{K}}|\mathbf{h}_k^H\mathbf{p}_{\mathrm{p},q}|^2 + \sigma_z^2}, 
\end{equation}
\begin{equation}
  \begin{aligned}
    % \gamma_{\mathrm{c}1,k} &= \frac{|\mathbf{h}_k^H\mathbf{w}_{\mathrm{c}1}|^2}{\sum_{g \in \mathcal{G}}|\mathbf{h}_k^H\mathbf{w}_{\mathrm{c}2,g}|^2 + \sum_{q \in \mathcal{K}}|\mathbf{h}_k^H\mathbf{w}_{\mathrm{p},q}|^2 + \sigma_k^2}, \forall k,\\
    % \gamma_{\mathrm{c}2,g,k} &= \frac{|\mathbf{h}_k^H\mathbf{w}_{\mathrm{c}2,g}|^2}{\sum_{\substack{o \in \mathcal{G}\\ o \ne g}}|\mathbf{h}_k^H\mathbf{w}_{\mathrm{c}2,o}|^2 + \sum_{q \in \mathcal{K}}|\mathbf{h}_k^H\mathbf{w}_{\mathrm{p},q}|^2 + \sigma_k^2}, \forall g, \forall k,\\
    \gamma_{\mathrm{p},k}= &\frac{|\mathbf{h}_k^H\mathbf{p}_{\mathrm{p},k}|^2}{\sum_{\substack{o \in \mathcal{G}\\ o \ne g}}|\mathbf{h}_k^H\mathbf{p}_{\mathrm{c}2,o}|^2 + \sum_{\substack{q \in \mathcal{K}\\ q \ne k}}|\mathbf{h}_k^H\mathbf{p}_{\mathrm{p},q}|^2 + \sigma_z^2}, \\
    &~~~~~~~~~~~~~~~~~~~~~~~~~~~\forall k \in \mathcal{K}, \forall g \in \mathcal{G}.
  \end{aligned}
\end{equation}

Similar to 1-layer RSMA, the inter- and inner-group common rate should satisfy $R_{\mathrm{c}1} = \min_{k \in \mathcal{K}} R_{\mathrm{c}1,k}$ and $R_{\mathrm{c}2,g} = \min_{k \in \mathcal{K}_g} R_{\mathrm{c}2,g,k}$, $\forall g \in \mathcal{G}$ to ensure successful decoding of the corresponding common streams. 

\subsection{Architecture for RIS} 

Based on the analysis in \cite{Shanpu Shen}, an $M$-element RIS can be equivalent to an $M$-port network, where a reconfigurable impedance network is constructed to adaptively reflect signals and shape propagation environments. According to the circuit connection topology, RISs can be classified into the following three categories.

\subsubsection{Single-connected RIS} In this category, each port (equivalently each RIS element) is connected to ground via a reconfigurable impedance without any connections with other ports. RIS coefficients are thus diagonal matrices and each non-zero element has a constant magnitude, i.e., 
\begin{equation}
  \mathbf{\Phi}_l = \mathsf{diag}(\phi_{1,l}, \ldots, \phi_{M,l}),|\phi_{m,l}| = 1, \forall m \in \mathcal{M}, \forall l \in \mathcal{L}.
\end{equation}

\subsubsection{Fully-connected RIS} In \cite{Shanpu Shen}, the authors propose a more general RIS model, namely, fully-connected RIS. Different from single-connected RIS, each port in fully-connected RIS is connected to all other ports via reconfigurable impedances. Therefore, RIS coefficients are full matrices satisfying 
\begin{equation}
  \mathbf{\Phi}_l^H\mathbf{\Phi}_l = \mathbf{I}_M, \mathbf{\Phi}_l = \mathbf{\Phi}_l^T, \forall l \in \mathcal{L}. 
\end{equation}

\subsubsection{Group-connected RIS} Fully-connected RIS is able to perform better and more flexible beam controlling than single-connected RIS, but at the expense of high hardware complexity. To finely balance the beam controlling accuracy and hardware complexity, the authors in \cite{Shanpu Shen} propose a group-connected RIS architecture. The main idea is to divide all $M$ elements into $S$ groups indexed by $\mathcal{S} = \{1, \ldots, S\}$. Each group contains $M_s$ elements, $\sum_{s \in \mathcal{S}} M_s = M$, and forms a small-dimensional fully-connected network. Denote $\mathcal{M}_s$ as the subset of RIS element indices for group $s$, $\forall s \in \mathcal{S}$. Then RIS coefficients can be modeled as block diagonal matrices, and each block is a full matrix, i.e.,
\begin{equation}
  \begin{aligned}
  \mathbf{\Phi}_l  &= \mathsf{blkdiag}(\mathbf{\Phi}_{\mathcal{M}_1,l}, \ldots, \mathbf{\Phi}_{\mathcal{M}_S,l}), \\
  \mathbf{\Phi}_{\mathcal{M}_s,l}^H\mathbf{\Phi}_{\mathcal{M}_s,l} &= \mathbf{I}_{M_s},
  \mathbf{\Phi}_{\mathcal{M}_s,l} = \mathbf{\Phi}_{\mathcal{M}_s,l}^T, \forall s \in \mathcal{S}, \forall l \in \mathcal{L}.
  \end{aligned}
\end{equation}

\begin{remark}
It is worthwhile to highlight here that the multi-RIS-aided multi-AP MU-MISO transmission network described in this section can be regarded as a general system model including all the existing RIS-aided RSMA schemes. On one hand, when the transmission network adopts a 1-layer RSMA, the illustrated general model boils down to \cite{Z Yang}, \cite{A Bansal} with the number of APs $N = 1$, to \cite{K Weinberger}, \cite{K Weinberger Journal} with the number of RISs $L = 1$, and to \cite{H Fu}, \cite{T Fang} with both $N = 1$ and $L = 1$; when the transmission network applies a 2-layer RSMA and $N = 1$, the illustrated model boils down to the model utilized in \cite{A Bansal Open J}, \cite{A Jolly}. This system model can also be further extended by embracing the generalized RSMA. On the other hand, most of the existing work \cite{Z Yang}-\cite{H Fu}, \cite{A Bansal Open J}, \cite{A Jolly} considers the deployment of single-connected RISs; only \cite{T Fang} considers designs of both single- and fully-connected RISs. 
\end{remark}

\begin{figure}
  \centering
  \subfigure[Perfect CSI]{
  \includegraphics[width=1.6 in]{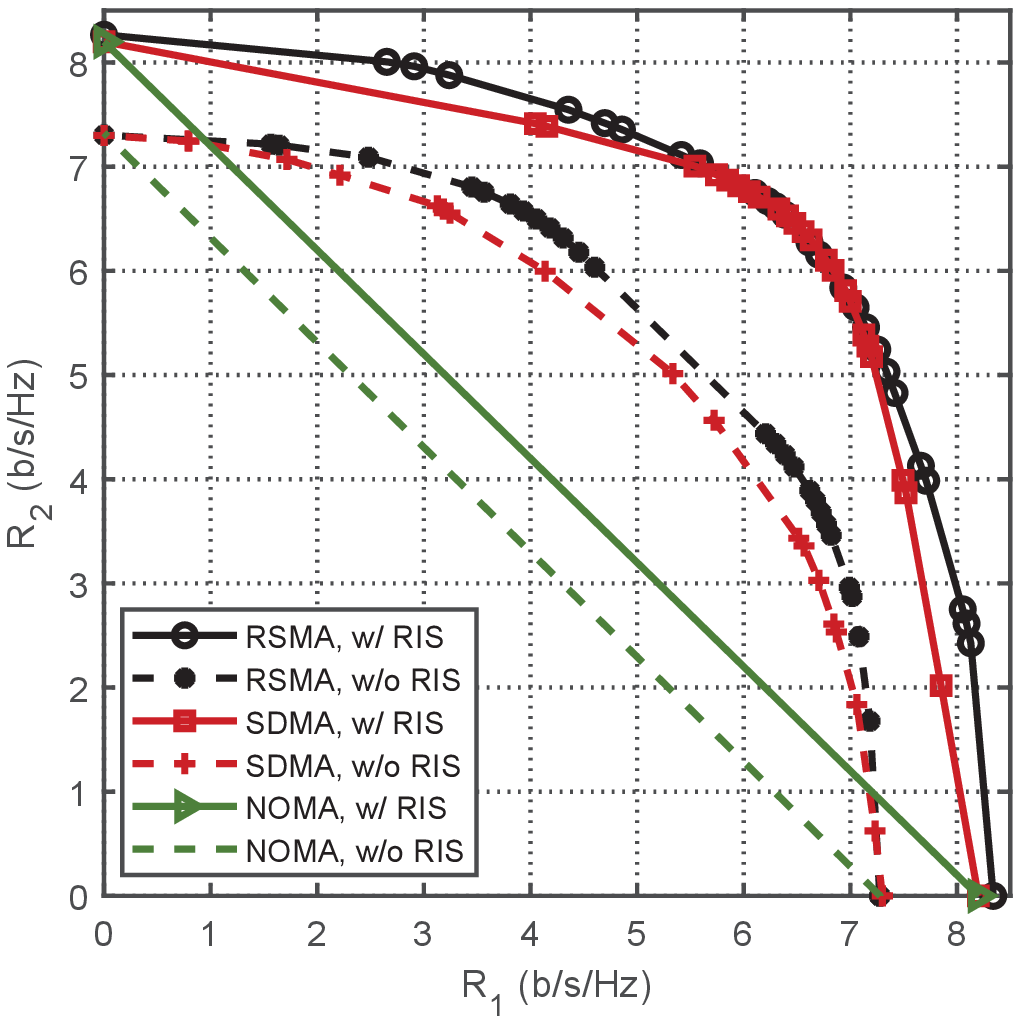}
  \label{fig:perfect_csi}}
  \subfigure[Imperfect CSI]{
  \includegraphics[width=1.6 in]{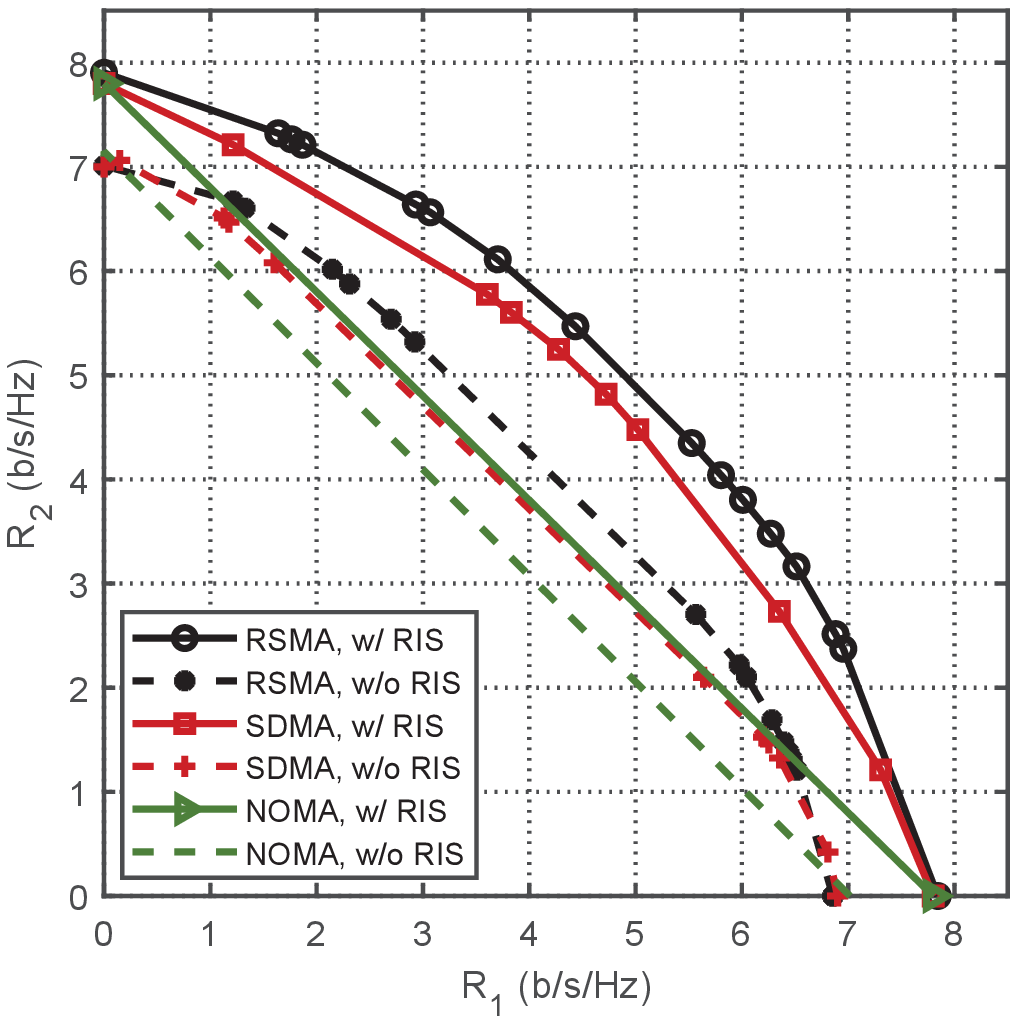}
  \label{fig:imperfect_csi}}
  \caption{Rate region comparison for a RIS-aided MU-MISO system with different CSI conditions ($N = 1$, $N_\mathrm{t} = K = 2$, $M = 20$, $P = 1$W). The AP applies a 1-layer RSMA and the RIS is assumed to have a single-connected architecture. Channels for direct links from the AP to users are realized by Rayleigh fading models to account for small-scale fading. Channels from the AP to the RIS, and from the RIS to users are modeled as Rician fading with Rician factor $\kappa = 2$ dB. The large-scale fading is modeled as $\kappa_i = \sqrt{\zeta_0 d_i^{-\epsilon_i}}$, $\forall i \in \{\mathrm{au, ar, ru}\}$, with reference signal attenuation (1 m) $\zeta_0 = -30$ dB, where subscripts ``$\mathrm{au, ar, ru}$'' refer to the AP-user link, the AP-RIS link, and the RIS-user link, respectively. Path loss exponents are set as $\epsilon_\mathrm{au} = 3$, $\epsilon_\mathrm{ar} = 1.9$, and $\epsilon_\mathrm{ru} = 1.7$, respectively. Distances between the AP and RIS, and between the RIS and users are $d_\mathrm{ar} = 50$ m and  $d_\mathrm{ru} = 10$ m, respectively. The distance between the AP and two users are set as $d_\mathrm{bu} = \sqrt{d_\mathrm{ar}^2 - d_\mathrm{ru}^2}$. The noise power at each user is $\sigma_z^2 = -70$ dBm.}
  \label{fig:rate_region}\vspace{-0.3 cm}
\end{figure}

\begin{remark}
Given the above RIS-aided and RSMA-enabled system model, existing works have investigated joint transmit beamforming and phase design by considering various metrics, which include energy efficiency maximization \cite{Z Yang}, \cite{K Weinberger Journal}, outage probability analysis \cite{A Bansal}, \cite{A Bansal Open J}, power minimization \cite{K Weinberger}, max-min fairness \cite{H Fu}, and sum-rate maximization \cite{T Fang}, \cite{A Jolly}.
However, the above-mentioned works are limited to beamforming designs for RIS-aided RSMA networks. The research beyond beamforming will be discussed in Section IV.  
\end{remark}

\section{Advantages of RIS-aided RSMA Networks}

In this section, we will discuss advantages of integrating RSMA with RIS from various perspectives and provide some simulation results to support our statements.

\vspace{-0.3 cm}

\subsection{Higher Spectral Efficiency}

It was proved in existing research that the spectral efficiency of RSMA is always higher than or at least equal to other multiple access techniques regardless of channel conditions \cite{Lina Mao}, \cite{Lina Mao 2018}, \cite{Anup}, \cite{Onur Dizdar}. Recent research on RIS-aided RSMA has also indicated the spectral efficiency enhancement of RIS-aided RSMA compared to RIS-aided SDMA/NOMA \cite{T Fang}, \cite{A Jolly}. 
In Fig. \ref{fig:rate_region}, we plot the rate regions of RSMA/SDMA/NOMA with and without the aid of RIS under different channel conditions\footnote{The joint transmit beamformer and phase design of Fig. \ref{fig:rate_region} is based on the equivalence between sum-rate maximization and mean square error minimization similar to \cite{Lina Mao 2018}. Specifically, the optimization for transmit precoder is a convex problem, which can be solved by interior-point methods \cite{Lina Mao 2018}. Based on the RIS modeling in \cite{Shanpu Shen}, the optimization for RIS can be transformed into an unconstrained problem, which can be solved by quasi-Newton methods. Although we consider the simplest case with only 2 users and one RIS, the algorithm can be easily extended to multi-RIS and multi-user cases with simple scales of matrix dimensions.}. It can be observed from Fig. \ref{fig:rate_region} that rate regions of three multiple access techniques are all enlarged with the aid of RIS. In addition, the achievable rate region of the RIS-aided RSMA scheme is greater than that of RIS-aided SDMA and NOMA, which again, demonstrates that integrating RSMA with RIS can achieve higher spectral efficiency.

% \subsubsection{Enhancement on Other Performance Metrics} Besides spectral efficiency, existing work has also shown the effectiveness of the integration of RSMA and RIS on maximizing energy efficiency \cite{Z Yang}, \cite{K Weinberger Journal}, minimizing outage probability \cite{A Bansal}, \cite{A Bansal Open J}, minimizing transmit power \cite{K Weinberger}, and performing better user fairness \cite{H Fu} compared to other RIS-aided schemes, e.g., NOMA, orthogonal frequency-diversion multiple access (OFDMA). 

\vspace{-0.3 cm}

\subsection{Coverage Extension and Beam Control Flexibility}

RSMA builds a flexible ``bridge'' between SDMA and NOMA. By adaptively adjusting the power ratio between common and private streams, RSMA is able to achieve smart interference management, and thus, is widely suitable for different scenarios, e.g., overloaded/underloaded cases and different user deployments. On the other hand, practical wireless communication systems inevitably suffer blockages, e.g., due to high buildings, which, to some extent, limits the communication coverage. RIS can be deployed as another ``bridge'', i.e., an additional path, between the transmitter and the receiver to finely address this issue. By combining those two ``bridges'', the communication coverage can be physically extended and the beam control flexibility can be further enhanced with the joint consideration of RSMA and RIS beamformer design.  

\vspace{-0.3 cm}

\subsection{Robust to CSI and SIC Imperfections, and User Mobility}

Accurate and instantaneous channel acquisition for AP-RIS-user link is difficult since there is no RF chain at the RIS. In the meanwhile, RSMA has been shown to be robust to CSI inaccuracies \cite{Anup} and user mobility \cite{Onur Dizdar}, and achieve low-latency communications \cite{Onur Dizdar}. Therefore, the interplay between RSMA and RIS is a potential new paradigm to compensate for the limitations of RIS and to fully utilize the benefits of both techniques to enhance system performance. In Fig. \ref{fig:imperfect_csi} we illustrate the rate region of RIS-aided RSMA/SDMA/NOMA under imperfect CSI\footnote{The channel estimation errors for AP-user, AP-RIS, and RIS-user channels for imperfect CSI conditions are modeled as statistical complex Gaussian entries in accordance with \cite{Anup}. The power of channel estimation error is set as $\sigma_e^2 = \sigma_z^2(1-\alpha)$ with $\alpha = 0.9$.}. It can be shown in Fig. \ref{fig:imperfect_csi} that RIS-aided RSMA still achieves a larger rate region than SDMA and NOMA.
More importantly, the rate region gap between (RIS-aided) RSMA and (RIS-aided) SDMA increases in imperfect CSI conditions, which demonstrates the robustness of RIS-aided RSMA to channel imperfections.
SIC imperfection in RSMA due to error propagation, channel estimation errors, and other impairments are also important factors which can cause performance loss. Fortunately, the recently published results in \cite{A SS} have shown the robustness of RIS-aided RSMA to SIC errors thanks to the assistance of RIS and flexible splitting between common and private messages.

\vspace{-0.3 cm}

\subsection{Lower Computational and Hardware Complexity}

Thanks to the newly introduced reflect link with the aid of RIS in wireless communication systems, it is possible to employ the simplest 1-layer RSMA scheme to achieve similar performance to (or even better performance than) multi-layer RSMA without RIS \cite{A Bansal}. It was shown in \cite{Lina Mao} that 1-layer RSMA outperforms existing multiple access schemes, e.g., multi-antenna NOMA, in terms of sum-rate but with lower receiver complexity. Moreover, Fig. \ref{fig:rate_region} demonstrates the advantage of RIS-aided 1-layer RSMA compared to RIS-aided NOMA. 
The abovementioned results indicate that integrating RSMA with RIS helps to reduce hardware and computational complexity of RSMA.  
Recent work in \cite{T Fang} shows that 1-layer RSMA with single-connected RIS can achieves almost the same (or even better) performance than SDMA with fully-connected RIS.  
This observation highlights the advantage of integrating RSMA with RIS on reducing RIS hardware complexity and computational complexity of the phase design. 

\vspace{-0.1 cm}

\section{Conclusion and Open Problems}

In this letter, we provide an overview of the integration of two state-of-the-art techniques: RSMA and RIS. We first introduce the principle of these two techniques and explain the motivation/need for integrating RSMA with RIS. We develop a general RIS-aided RSMA model which could include all of the existing works on RIS-aided RSMA, and compare different RSMA/RIS architectures. We then summarize the major advantages of RIS-aided RSMA networks, and for the first time illustrate the rate region comparison of RIS-aided RSMA, NOMA, and SDMA for both perfect and imperfect CSI settings.   
The study of RIS-aided RSMA is still in its infancy. There are many open problems worth being studied in the near future, which include, but are not limited to:

\subsubsection{RIS-Aided Uplink RSMA} 

 Uplink RSMA is deemed as an efficient way to achieve the capacity region of Gaussian multiple access channels without time sharing, which induces heavy overhead and synchronization issues, among users \cite{Lina Mao}. All of the existing research on RIS-aided RSMA \cite{Z Yang}-\cite{A Jolly} is restricted to downlink RSMA. When it comes to RIS-aided uplink RSMA, research issues such as beamforming design, and RIS deployments, e.g., how to choose the number and locations of RISs, are worth being considered and investigated.

\subsubsection{The Impact of Different RIS Constraints and RIS Deployments on RIS-Aided RSMA}
Existing research on RIS-aided RSMA mainly focused on joint transmit beamforming and single-connected RIS design \cite{Z Yang}-\cite{H Fu}, \cite{A Bansal Open J}, \cite{A Jolly} and only the recent work \cite{T Fang} considered both single- and fully-connected RIS design. It is interesting to consider the joint RSMA and RIS design when RIS has a group-connected architecture. In addition, optimization for practical RIS with discrete values is another important and meaningful topic. Quantization for single-connected RIS is straightforward and has been considered in existing research on RIS. However, when it comes to group/fully-connected cases, how to quantize RIS coefficients is not an easy problem. Existing work in \cite{Matteo} has investigated group/fully-connected RIS with discrete values but in a simple single-input single-output (SISO) system. It is interesting to consider the joint optimization of transmit precoder and RIS beamformer with discrete values for RIS-aided RSMA networks. 
The deployment of RIS is also important and has been investigated in RIS-aided SDMA systems. It would be meaningful to study the impact of RIS deployments in RIS-aided RSMA networks.

\subsubsection{Interplay With Other Techniques} 
Previous works are limited to the integration of RIS and RSMA in wireless communication systems \cite{Z Yang}-\cite{A Jolly}. In the future, it is interesting to integrate RSMA, RIS, as well as other techniques, e.g., sensing/radar detection, WPT, visible light communication, to establish multi-functional systems, which highly matches with one of the most important aims of future networks: develop smart and unified systems that can access any standard and provide seamless service.

\end{document}